\documentclass[12pt]{article}
\pdfoutput=1
\usepackage[colorlinks,linkcolor=Blue,citecolor=Blue,bookmarks,bookmarksnumbered]{hyperref}
\usepackage[scaled=0.85]{helvet}
\usepackage{amsmath,amssymb,accents,mathrsfs,XoohmE}
\usepackage{graphicx,color}
\graphicspath{{Figures/}}
\usepackage{booktabs}
\usepackage{multirow}
\usepackage{placeins}
\usepackage{amsmath}
\usepackage{subfigure}

\usepackage{XoohmE}

\definecolor{Green}  {rgb}{0.10,0.70,0.10} 
\definecolor{Orange} {rgb}{1.00,0.50,0.15} 
\definecolor{Red}    {rgb}{0.90,0.00,0.12} 
\definecolor{Purple} {rgb}{0.50,0.25,0.55} 
\definecolor{Turque} {rgb}{0.00,0.65,0.85} 
\definecolor{Blue}   {rgb}{0.00,0.00,1.00} 
\definecolor{Magenta}{rgb}{1.00,0.00,1.00} 
\definecolor{Gold}   {rgb}{1.00,0.75,0.25} 
\definecolor{Seaweed}{rgb}{0.01,0.24,0.09} 
\definecolor{Brown}  {rgb}{0.43,0.26,0.32} 
\definecolor{grey1}  {rgb}{0.20,0.20,0.20} 
\definecolor{grey2}  {rgb}{0.40,0.40,0.40} 
\definecolor{grey3}  {rgb}{0.60,0.60,0.60} 
\definecolor{grey4}  {rgb}{0.80,0.80,0.80} 
\definecolor{grey5}  {rgb}{0.90,0.90,0.90} 
\def\C#1#2{{\ifcase#1\or
             \color{Green}\or \color{Orange}\or \color{Red}\or
              \color{Purple}\or \color{Turque}\or \color{Blue}\or
               \color{Magenta}\or \color{Gold}\or \color{Seaweed}\or
                \color{Brown}\or\color{grey1}\or\color{grey2}\or
                 \color{grey3}\else\color{grey4}\fi#2}}

\definecolor{Slate} {rgb}{0.00,0.45,0.55}

\def\rI{{\rm I}}
\def\rJ{{\rm J}}

\def\rL{{\rm L}}
\def\rR{{\rm R}}

\def\fracm#1#2{\hbox{\large{${\frac{{#1}}{{#2}}}$}}}

\def\be{\begin{equation}}
\def\ee{\end{equation}}
\newcommand{\bea}{\begin{eqnarray}}
\newcommand{\eea}{\end{eqnarray}}
\newcommand{\ena}{\end{eqnarray}}

\def\pp{{\mathchoice
          {
              \kern 1pt%
              \raise 1pt
              \vbox{\hrule width5pt height0.4pt depth0pt
                    \kern -2pt
                    \hbox{\kern 2.3pt
                          \vrule width0.4pt height6pt depth0pt
                          }
                    \kern -2pt
                    \hrule width5pt height0.4pt depth0pt}%
                    \kern 1pt
           }
            {
              \kern 1pt%
              \raise 1pt
              \vbox{\hrule width4.3pt height0.4pt depth0pt
                    \kern -1.8pt
                    \hbox{\kern 1.95pt
                          \vrule width0.4pt height5.4pt depth0pt
                          }
                    \kern -1.8pt
                    \hrule width4.3pt height0.4pt depth0pt}%
                    \kern 1pt
            }
            {
              \kern 0.5pt%
              \raise 1pt
              \vbox{\hrule width4.0pt height0.3pt depth0pt
                    \kern -1.9pt  
                    \hbox{\kern 1.85pt
                          \vrule width0.3pt height5.7pt depth0pt
                          }
                    \kern -1.9pt
                    \hrule width4.0pt height0.3pt depth0pt}%
                    \kern 0.5pt
            }
            {
              \kern 0.5pt%
              \raise 1pt
              \vbox{\hrule width3.6pt height0.3pt depth0pt
                    \kern -1.5pt
                    \hbox{\kern 1.65pt
                          \vrule width0.3pt height4.5pt depth0pt
                          }
                    \kern -1.5pt
                    \hrule width3.6pt height0.3pt depth0pt}%
                    \kern 0.5pt
            }
        }}

\def\mm{{\mathchoice
                       {
                             \kern 1pt
               \raise 1pt    \vbox{\hrule width5pt height0.4pt depth0pt
                                  \kern 2pt
                                  \hrule width5pt height0.4pt depth0pt}
                             \kern 1pt}
                       {
                            \kern 1pt
               \raise 1pt \vbox{\hrule width4.3pt height0.4pt depth0pt
                                  \kern 1.8pt
                                  \hrule width4.3pt height0.4pt depth0pt}
                             \kern 1pt}
                       {
                            \kern 0.5pt
               \raise 1pt
                            \vbox{\hrule width4.0pt height0.3pt depth0pt
                                  \kern 1.9pt
                                  \hrule width4.0pt height0.3pt depth0pt}
                            \kern 1pt}
                       {
                           \kern 0.5pt
             \raise 1pt  \vbox{\hrule width3.6pt height0.3pt depth0pt
                                  \kern 1.5pt
                                  \hrule width3.6pt height0.3pt depth0pt}
                           \kern 0.5pt}
                       }}

\def\ad{{\kern0.5pt
                   \alpha \kern-5.05pt \raise5.8pt\hbox{$\textstyle.$}\kern
0.5pt}}

\def\bd{{\kern0.5pt
                   \beta \kern-5.05pt \raise5.8pt\hbox{$\textstyle.$}\kern
0.5pt}}

\def\qd{{\kern0.5pt
                   q \kern-5.05pt \raise5.8pt\hbox{$\textstyle.$}\kern
0.5pt}}
\def\Dot#1{{\kern0.5pt
     {#1} \kern-5.05pt \raise5.8pt\hbox{$\textstyle.$}\kern
0.5pt}}

\catcode`@=11
\def\un#1{\relax\ifmmode\@@underline#1\else
        $\@@underline{\hbox{#1}}$\relax\fi}
\catcode`@=12

\def\a{\alpha}
\def\b{\beta}
\def\c{\chi}
\def\d{\delta}
\def\e{\epsilon}

\def\g{\gamma}

\def\i{\iota}
\def\j{\psi}
\def\k{\kappa}
\def\l{\lambda}
\def\m{\mu}
\def\n{\nu}
\def\o{\omega}

\def\r{\rho}
\def\s{\sigma}
\def\t{\tau}

\def\L{\Lambda}
\def\O{\Omega}
\def\P{\Pi}

\def\S{\Sigma}

\def\ca{{\cal A}}

\def\cf{{\cal F}}

\def\dslash{\not{\hbox{\kern-2pt $\partial$}}}
\def\Dslash{\not{\hbox{\kern-4pt $D$}}}
\def\pslash{\not{\hbox{\kern-2.3pt $p$}}}
 \newtoks\slashfraction
 \slashfraction={.13}
 \def\slash#1{\setbox0\hbox{$ #1 $}
 \setbox0\hbox to \the\slashfraction\wd0{\hss \box0}/\box0 }
 
\def\kcr{{\hbox{\ro \char'170}}}                
\def\ktl{{\hbox{\ro \char'170}}}        
\def\ktr{{\hbox{\ro \char'170}}}        
\def\kbl{{\hbox{\ro \char'170}}}        
\def\kbr{{\hbox{\ro \char'170}}}        

\def\plpl{\raise-2pt\hbox{$\raise3pt\hbox{$_+$}\hskip-6.67pt\raise0.0pt
\hbox{$^+$}\hskip 0.01pt$}}
\def\mimi{\raise-2pt\hbox{$\raise3pt\hbox{$_-$}\hskip-6.67pt\raise0.0pt
\hbox{$^-$}\hskip 0.01pt$}} 

\def\bo{{\raise.15ex\hbox{\large$\Box$}}}               
\def\pa{\partial}                                       
\def\TH{{\raise.2ex\hbox{$\displaystyle \bigodot$}\mskip-4.7mu \llap H \;}}
\def\face{{\raise.2ex\hbox{$\displaystyle \bigodot$}\mskip-2.2mu \llap {$\ddot
        \smile$}}}                                      

\def\dt#1{\on{\hbox{\bf .}}{#1}}                
\def\Dot#1{\dt{#1}}

\def\Hat#1{\widehat{#1}}                        
\def\leftrightarrowfill{$\mathsurround=0pt \mathord\leftarrow \mkern-6mu
        \cleaders\hbox{$\mkern-2mu \mathord- \mkern-2mu$}\hfill
        \mkern-6mu \mathord\rightarrow$}
\def\dvec#1{\vbox{\ialign{##\crcr
        \leftrightarrowfill\crcr\noalign{\kern-1pt\nointerlineskip}
        $\hfil\displaystyle{#1}\hfil$\crcr}}}           
\def\dt#1{{\buildrel {\hbox{\LARGE .}} \over {#1}}}     

\def\fracm#1#2{\hbox{\large{${\frac{{#1}}{{#2}}}$}}}
\def\sfrac#1#2{{\vphantom1\smash{\lower.5ex\hbox{\small$#1$}}\over
        \vphantom1\smash{\raise.4ex\hbox{\small$#2$}}}} 
\def\bfrac#1#2{{\vphantom1\smash{\lower.5ex\hbox{$#1$}}\over
        \vphantom1\smash{\raise.3ex\hbox{$#2$}}}}       
\def\afrac#1#2{{\vphantom1\smash{\lower.5ex\hbox{$#1$}}\over#2}}    

\def\pa{\partial}      
\let\bm\relax
\newcommand{\bm}[1]{{\boldsymbol{#1}}}

\def\ad{{\dot{\alpha}}}
\def\bd{{\dot{\beta}}}

 \font\rOpe=cmsy10                        
 \def\ktl{{\hbox{\rOpe\char'170}}}        
 \def\kbl{{\hbox{\rOpe\char'170}}}        
 \def\kcr{{\reflectbox{\rOpe\char'170}}}        
 \def\ktr{{\reflectbox{\rOpe\char'170}}}        
 \def\kbr{{\reflectbox{\rOpe\char'170}}}        
 \def\Border{\vbox{\hsize0pt
        \setlength{\unitlength}{1mm}
        \newcount\xco
        \newcount\yco
        \xco=-21
        \yco=12
        \begin{picture}(0,0)(-7.5,0)
        \put(\xco,\yco){$\ktl$}
        \advance\yco by-1
        {\loop
        \put(\xco,\yco){$\kcr$}
        \advance\yco by-2
        \ifnum\yco>-240
        \repeat
        \put(\xco,\yco){$\kbl$}}
        \xco=170
        \yco=12
        \put(\xco,\yco){$\ktr$}
        \advance\yco by-1
        {\loop
        \put(\xco,\yco){$\kcr$}
        \advance\yco by-2
        \ifnum\yco>-240
        \repeat
        \put(\xco,\yco){$\kbr$}}
        \put(-19.5,13){\scalebox{.6065}{%
         University of Maryland Center for String and Particle  Theory \&\ Physics Department%
        |University of Maryland Center for String and Particle  Theory \&\ Physics Department}}
        \put(-19.5,-241.5){\scalebox{.5835}{%
         ****University of Maryland * Center for String and
         Particle  Theory* Physics Department****University of Maryland *Center
        for String and Particle  Theory* Physics Department}}
        \end{picture}
        \par\vskip-8mm}}
\definecolor{UMred}{rgb}{.9,.05,.2}
\definecolor{HUblue}{rgb}{.0,.3,.7}

\definecolor{Red}    {rgb}{0.90,0.00,0.12} 
\definecolor{Blue}   {rgb}{0.00,0.00,1.00} 
\definecolor{Green}  {rgb}{0.10,0.70,0.10} 
\definecolor{Turque} {rgb}{0.00,0.65,0.85} 
\definecolor{Orange} {rgb}{1.00,0.50,0.15} 
\definecolor{Magenta}{rgb}{1.00,0.00,1.00} 
\definecolor{Gold}   {rgb}{1.00,0.75,0.25} 
\definecolor{Seaweed}{rgb}{0.01,0.24,0.09} 
\definecolor{Purple} {rgb}{0.50,0.25,0.55} 
\definecolor{Brown}  {rgb}{0.43,0.26,0.32} 
\definecolor{grey1}  {rgb}{0.20,0.20,0.20} 
\definecolor{grey2}  {rgb}{0.40,0.40,0.40} 
\definecolor{grey3}  {rgb}{0.60,0.60,0.60} 
\definecolor{grey4}  {rgb}{0.80,0.80,0.80} 
\definecolor{grey5}  {rgb}{0.90,0.90,0.90} 
\def\C#1#2{{\ifcase#1\or
             \color{Red}\or \color{Green}\or \color{Blue}\or\
              \color{Turque}\or \color{Orange}\or \color{Magenta}\or 
               \color{Gold}\or \color{Seaweed}\or \color{Purple}\or
                \color{Brown}\or\color{grey1}\or\color{grey2}\or
                 \color{grey3}\else\color{grey4}\fi#2}}

\definecolor{Slate} {rgb}{0.00,0.45,0.55}

\newdimen\parshift\parshift=\parindent
\catcode`@=11
 \long\def\@footnotetext#1{\insert\footins{\reset@font\footnotesize
           \interlinepenalty\interfootnotelinepenalty\splittopskip%
            \footnotesep\splitmaxdepth\dp\strutbox\floatingpenalty\@MM%
             \hsize\columnwidth\addtolength{\hsize}{-2\parindent}
              \@parboxrestore\protected@edef\@currentlabel%
              {\csname p@footnote\endcsname\@thefnmark}%
                \color@begingroup%
                 \@makefntext{\rule\z@\footnotesep\ignorespaces#1%
                  \@finalstrut\strutbox}%
                \color@endgroup}}
 \long\def\@makefntext#1{\hglue\parshift%
           \vbox{\noindent\baselineskip=11pt plus.5pt minus.5pt\hb@xt@0em{\hss\@makefnmark\kern1pt}#1}}
\catcode`@=12

\newskip\humongous \humongous=0pt plus 1000pt minus 1000pt
\def\caja{\mathsurround=0pt}
\def\eqalign#1{\,\vcenter{\openup2\jot \caja
        \ialign{\strut \hfil$\displaystyle{##}$&$
        \displaystyle{{}##}$\hfil\crcr#1\crcr}}\,}
\newif\ifdtup
\def\panorama{\global\dtuptrue \openup2\jot \caja
        \everycr{\noalign{\ifdtup \global\dtupfalse
        \vskip-\lineskiplimit \vskip\normallineskiplimit
        \else \penalty\interdisplaylinepenalty \fi}}}

\makeatletter
\def\section{\@startsection{section}{1}{\z@}
        {3ex plus-1ex minus-.2ex}{1pt plus1pt}{\large\sf\bfseries\boldmath}}
\def\subsection{\@startsection{subsection}{2}{\z@}
         {1.5ex plus-1ex minus-.2ex}{0.01pt plus1pt}{\sf\slshape}}
\def\subsubsection{\@startsection{subsubsection}{3}{\z@}
          {1.5ex plus-1ex minus-.2ex}{0.01pt plus0.2pt}{\sf\boldmath}}
\def\paragraph{\@startsection{paragraph}{4}{\z@}
           {.75ex \@plus.5ex \@minus.2ex}{-2mm}{\sf\bfseries\boldmath}}
\makeatother

 \allowdisplaybreaks
 \seceq

\usepackage{lipsum}
\usepackage{listings}
\definecolor{MyDarkGreen}{rgb}{0.0,0.4,0.0} 
\usepackage[enableskew,vcentermath]{youngtab}

\definecolor{Hey}{rgb}{.9,.05,.4}
\definecolor{orange}{rgb}{1,.5,0}
\definecolor{plum}{rgb}{.4,0,.6}
\definecolor{R}{rgb}{1,0,0}
\definecolor{G}{rgb}{0.1,0.7,0}
\definecolor{B}{rgb}{0,0,1}
\begin{document}

\thispagestyle{empty}
\noindent{\small
\hfill{  \\ 
$~~~~~~~~~~~~~~~~~~~~~~~~~~~~~~~~~~~~~~~~~~~~~~~~~~~~~~~~~~~~~~~~~$
$~~~~~~~~~~~~~~~~~~~~~~~~~~~~~~~~~~~~~~~~~~~~~~~~~~~~~~~~~~~~~~~~~$
{}
}}
\vspace*{0mm}
\begin{center}
{\large \bf
A Letter Highlighting Matrix Mapping in \vskip1pt
Minimal 4D, $\bm {\cal N}$ = 1  On-Shell Supermultiplet Representations
 \\[2pt]
}   \vskip0.3in
{\large {
Delilah E.\ A.\ Gates \footnote{dgates@g.harvard.edu}$^{,a}$, and
S.\ James Gates, Jr.\footnote{sylvester${}_-$gates@brown.edu}$^{,b,c}$
$~~~~~~$
}}
\\*[8mm]
\emph{
\centering
$^{a}$Center for the Fundamental Laws of Nature, Harvard University,
\\[1pt]
Cambridge, MA 02138, USA,
\\[4pt] and \\[4pt]
$^{b}$Brown University, Department of Physics,
\\[1pt]
Box 1843, 182 Hope Street, Barus \& Holley 545,
Providence, RI 02912, USA,
\\[4pt] and \\[4pt]
$^{c}$Brown Center for Theoretical Physics, 
340 Brook Street, Barus Hall,
Providence, RI 02912, USA
}
 \\*[80mm]
{ ABSTRACT}\\[4mm]
\parbox{142mm}{\parindent=2pc\indent\baselineskip=14pt plus1pt
On the basis of comparing eigenvalues of an operator ${\Hat {\bm {\cal C}}}{}^{({\cal R})}$,
that proved useful in distinguishing how off-shell 4D, $\cal N$ = 1 supermultiplets
become off-shell 4D, $\cal N$ = 2 supermultiplets, the double tensor supermultiplet
is shown to be radically different for other known multiplets.  This suggests
difficulties, if not impossibilities, to embed it into an off-shell structure.}
 \end{center}
\vfill
\noindent PACS: 11.30.Pb, 12.60.Jv\\
Keywords: adinkra, supersymmetry
\vfill
\clearpage
%

%
\section{Introduction}
\label{sec:NTR0}

It is the purpose of this letter to introduce into the literature approaches that allow on-shell 
supersymmetrical representations to be mapped (usually) into representations of Coxeter 
Groups \cite{Cx[1],Cx[2],Cx[3]}.  In a previous work \cite{permutadnk} such a mapping 
operation was defined for off-shell minimal 4D, $\cal N$ = 1 supersymmetrical multiplets.  
This allowed the super charges that act on supermultiplets to be mapped into Coxeter 
group elements.  An enabling feature of this mapping operator is that the mapping of a 
supercharge is determined by the off-shell supermultiplet on which the supercharge acts.

The important lessons based on this work as well as that in \cite{adnkKyeoh,EX2}: starting with SUSY transformation laws  generated by requiring invariances of on-shell actions (i.\ e.\ 
without auxiliary fields), typically leads to non-square  ${\bm {\rL}}_\rI^{({\cal R})}$ matrices. 
Non-square matrices can never describe the hidden Euclidean Clifford Algebras required for 
off-shell realizations of SUSY.  A necessary but not sufficient condition for realizing higher 
extended SUSY is that some of the $ {\bm {\rL}}_\rI^{({\cal R})}$ matrices must be square. By 
judicious choices of constructing larger square matrices from   smaller square matrices, higher
 extended off-shell SUSY's can be realized.
 
 We begin our presentation by reviewing results for the 4D, $\cal N$ = 1 Chiral, Tensor,
Vector, and Double Tensor supermultiplets found in the work of \cite{G-1}.

\subsection{On-Shell 4D, $\cal N$ = 1 Chiral Supermultiplet}

  We start
with chiral supermultiplet $(CM)$ where the supercovariant derivative (equivalent to the
supercharge) acts according to
\be \eqalign{
{\rm D}_a A ~&=~ \psi_a  ~~~, \cr
{\rm D}_a B ~&=~ i \, (\gamma^5){}_a{}^b \, \psi_b  ~~~, \cr
{\rm D}_a \psi_b ~&=~ i\, (\gamma^\mu){}_{a \,b}\,  \partial_\mu A 
~-~  (\gamma^5\gamma^\mu){}_{a \,b} \, \partial_\mu B   ~~~.
} \label{chi3}
\ee
After a series of calculation one finds the following results
\be \eqalign{  {~~~~~}
\{ ~ {\rm D}_a  \,,\,  {\rm D}_b ~\} \, A 
~&=~  i\, 2 \, (\gamma^\mu){}_{a \,b}\,  \partial_\mu \,  A ~~~, ~~~
\{ ~ {\rm D}_a  \,,\,  {\rm D}_b ~\} \, B 
~=~  i\, 2 \, (\gamma^\mu){}_{a \,b}\,  \partial_\mu \, B ~~~, \cr
\{ ~ {\rm D}_a  \,,\,  {\rm D}_b ~\} \, \psi{}_{c}  
~&=~  i\, 2 \, (\gamma^\mu){}_{a \,b}\,  \partial_\mu \,  \psi{}_{c}    
~-~   i \, (\gamma^\mu){}_{a \,b}\, (\gamma_\mu
\gamma^\nu){}_c{}^d  \partial_\nu \,  \psi{}_{d}     ~~~.
} \label{chi4}
\ee

\subsection{On-Shell 4D, $\cal N$ = 1 Tensor Supermultiplet}

Next, in the tensor supermultiplet $(TM)$ the supercovariant derivative acts according 
to
\be \eqalign{
{\rm D}_a \varphi ~&=~ \chi_a  ~~~, \cr
{\rm D}_a B{}_{\mu \, \nu } ~&=~ -\, \fracm 14 ( [\, \gamma_{\mu}
\, , \,  \gamma_{\nu} \,]){}_a{}^b \, \chi_b  ~~~, \cr
{\rm D}_a \chi_b ~&=~ i\, (\gamma^\mu){}_{a \,b}\,  \partial_\mu \varphi 
~-~  (\gamma^5\gamma^\mu){}_{a \,b} \, \e{}_{\mu}{}^{\r \, \s \, \t}
\partial_\r B {}_{\s \, \t}~~.
} \label{ten1}
\ee
Using this realization of the D-operator yields results of the form 
\be \eqalign{
\{ ~ {\rm D}_a  \,,\,  {\rm D}_b ~\} \, {\varphi}  
~&=~  i\, 2 \, (\gamma^\mu){}_{a \,b}\,  \partial_\mu \,  {\varphi}  ~~, \cr
{~~~~~~~}
\{ ~ {\rm D}_a  \,,\,  {\rm D}_b ~\} \, B{}_{\mu \, \nu }  
~&=~  i\, 2 \, (\gamma^\r){}_{a \,b}\,  \partial_\r \, B{}_{\mu \, \nu }  
~+~ \partial_\mu \, q{}_{\nu ~ a \, b} ~-~  \partial_\nu \, q{}_{\mu ~ 
a \, b} ~~, \cr
\{ ~ {\rm D}_a  \,,\,  {\rm D}_b ~\} \, \chi{}_{c}  
~&=~  i\, 2 \, (\gamma^\mu){}_{a \,b}\,  \partial_\mu \,  \chi{}_{c}    ~~,  ~~
q{}_{\mu ~ a \, b} ~\equiv~     i\, 2 \, (\gamma^\nu){}_{a \,b}   \, 
 [ \, B{}_{\mu \, \nu } \,+\, \fracm 12 \eta {}_{\mu \, \nu } \,\varphi \, ] ~~.
} \label{ten2}
\ee

\subsection{On-Shell 4D, $\cal N$ = 1 Vector Supermultiplet}

In the vector supermultiplet $(VM)$ the supercovariant derivative acts as
\be \eqalign{
{\rm D}_a \, A{}_{\mu} ~&=~  (\gamma^\mu){}_a {}^b \,  \l_b  ~~~, \cr
{\rm D}_a \l_b ~&=~   - \,i \, \fracm 14 ( [\, \gamma^{\mu}\, , \,  \gamma^{\nu} 
\,]){}_a{}_b \, (\,  \partial_\mu  \, A{}_{\nu} ~-~  \partial_\nu  \, A{}_{\mu} 
\, )  ~~~, \cr
} \label{V4}
\ee
and once again we calculate the anti-commutator 
to find
\be \eqalign{
{~~~~~~~}
\{ ~ {\rm D}_a  \,,\,  {\rm D}_b ~\} \, A{}_{\mu}  
~&=~  i\, 2 \, (\gamma^\r){}_{a \,b}\,  \partial_\r \, A{}_{\mu}  
~-~ \partial_\mu \, r{}_{a \, b}   ~~,  ~~
r{}_{a \, b} ~\equiv~     i\, 2 \, (\gamma^\nu){}_{a \,b}   \, A{}_{\nu }
~~,    \cr
\{ ~ {\rm D}_a  \,,\,  {\rm D}_b ~\} \,\l{}_c
~&=~  i\, 2 \, (\gamma^\mu){}_{a \,b}\,  \partial_\mu \,  \l{}_c   ~-~ i \,
\fracm 12 \,  (\gamma^\mu){}_{a \,b}\,  (\gamma_\mu  
\gamma^\nu){}_c{}^d \,   \partial_\nu \,  \l{}_d   \cr
&~~~~+~ i \, \fracm 1{16} \,  ([ \, \gamma^\a \, , \, \gamma^\b \,]){}_{a \,b}\,  
( [ \, \gamma_\a \, , \, \gamma_\b \,] \gamma^\nu){}_c{}^d \,   \partial_\nu \,  \l{}_d   
~~.
} \label{V5}
\ee
In both of the cases for the $(TM)$ and $(VM)$, when the Dirac equation is
imposed on the spinor, the algebra takes the form of the usual supersymmetry
definition.

\subsection{On-Shell 4D, $\cal N$ = 1 Double Tensor Supermultiplet}

In the double tensor supermultiplet $(DTM)$ the supercovariant derivative acts 
as
\be \eqalign{
{\rm D}_a X{}_{\mu \, \nu } ~&=~ i \, \fracm 14 (\gamma^5  [\, \gamma_{\mu}
\, , \,  \gamma_{\nu} \,]){}_a{}^b \, \L_b  ~~~, \cr
{\rm D}_a Y{}_{\mu \, \nu } ~&=~ - \, \fracm 14 ( [\, \gamma_{\mu}
\, , \,  \gamma_{\nu} \,]){}_a{}^b \, \L_b  ~~~, \cr
{\rm D}_a \L_b ~&=~ i\, (\gamma^\mu){}_{a \,b}\,  \e{}_{\mu}{}^{\r \, \s \, \t}
\partial_\r X {}_{\s \, \t} ~-~  (\gamma^5\gamma^\mu){}_{a \,b} \, 
\e{}_{\mu}{}^{\r \, \s \, \t} \partial_\r Y {}_{\s \, \t}~~,
} \label{dblten1}
\ee
and yields a very different commutation algebra
\be \eqalign{
{~~~~~~~}
\{ ~ {\rm D}_a  \,,\,  {\rm D}_b ~\} \, X{}_{\mu \, \nu }  
~&=~  i\, 2 \, (\gamma^\r){}_{a \,b}\,  \partial_\r \, X{}_{\mu \, \nu }  
~+~ \partial_\mu \, s{}_{\nu ~ a \, b} ~-~  \partial_\nu \, s{}_{\mu ~ 
a \, b}   \cr
&~~~~-~ i \, [~  \eta{}_{\a \, \mu } \, (\gamma_\nu){}_{a \,b} ~-~
\eta{}_{\a \, \nu } \, (\gamma_\mu){}_{a \,b}  ~]  \, \e{}^{\a \, \r \, \s \, \t}
\partial_\r Y {}_{\s \, \t}~~,    \cr
{~~~~~~~}
\{ ~ {\rm D}_a  \,,\,  {\rm D}_b ~\} \, Y{}_{\mu \, \nu }  
~&=~  i\, 2 \, (\gamma^\r){}_{a \,b}\,  \partial_\r \, Y{}_{\mu \, \nu }  
~+~ \partial_\mu \, t{}_{\nu ~ a \, b} ~-~  \partial_\nu \, t{}_{\mu ~
 a \, b}   \cr
&~~~~+~ i \, [~  \eta{}_{\a \, \mu } \, (\gamma_\nu){}_{a \,b} ~-~
\eta{}_{\a \, \nu } \, (\gamma_\mu){}_{a \,b}  ~]  \, \e{}^{\a \, \r \, \s \, \t}
\partial_\r X {}_{\s \, \t}~~,   \cr
 &s{}_{\mu ~ a \, b} ~\equiv~     i\, 2 \, (\gamma^\nu){}_{a \,b}   \, 
 X{}_{\mu \, \nu }   ~~,~~  t{}_{\mu ~ a \, b} ~\equiv~   i\, 2 \, 
 (\gamma^\nu){}_{a \,b} \, Y{}_{\mu \, \nu }  ~~,    \cr
{~~~~~~~}
\{ ~ {\rm D}_a  \,,\,  {\rm D}_b ~\} \, \L_c ~&=~ i\, 2\, (\gamma^\mu){}_{a \,b}\,  
\partial_\mu \,  \L_c ~+~ i \, (\gamma^\mu){}_{a \,b}\, (\gamma_\mu \, 
\gamma^\nu){}_c {}^d\,  \partial_\nu \,  \L_d    ~~~~.
}
\ee \label{dblten5}
The closure of the algebra on the bosons $X{}_{\m \n} $ and $Y{}_{\m \n} $
include the terms 
$$ \eqalign{
 &i \, [~  \eta{}_{\a \, \mu } \, (\gamma_\nu){}_{a \,b} ~-~
\eta{}_{\a \, \nu } \, (\gamma_\mu){}_{a \,b}  ~]  \, \e{}^{\a \, \r \, \s \, \t}
\partial_\r Y {}_{\s \, \t}  ~~~, \cr
& i \, [~  \eta{}_{\a \, \mu } \, (\gamma_\nu){}_{a \,b} ~-~
\eta{}_{\a \, \nu } \, (\gamma_\mu){}_{a \,b}  ~]  \, \e{}^{\a \, \r \, \s \, \t}
\partial_\r X {}_{\s \, \t}~~~,
}$$
respectively.  These may be interpreted as equations of motion
terms if there is an action that admits
\be \eqalign{
\d{}_Z ~=~ -  i 2\, \xi_{\m} \,   \e_{\n}{}^{\r \, \s \, \t} \left[ \,  (\partial_\r Y {}_{\s \, \t})
{ {\pa {~~~~}} \over {\pa X{}_{\m \n}} } ~-~  (\partial_\r X {}_{\s \, \t})
{ {\pa {~~~~}} \over {\pa Y{}_{\m \n}} } \,  \right] ~~,
}  \label{dblten7}
\ee
as the generator of a symmetry.

\section{From On-shell 4D, $\bm {\cal N} $ = 1 SUSY Multiplets to Non-Invertible
Matrices}
\label{sec:PeRMs}

Following the procedure in \cite{EX2} a definition
\be {
{\Hat {\bm \g}}{}_\rI^{({\cal R})}  ~=~ \frac12 \, (\, {\bm  \sigma}^1 \,+\, i {\bm  \sigma}^2  \, ) 
\, \otimes \,  {\bm {\rL}}_\rI^{({\cal R})}  ~+~
\frac12 \, (\, {\bm  \sigma}^1 \,-\, i {\bm  \sigma}^2  \, ) \, \otimes \,
[ \, {\bm {\rL}}_\rI^{({\cal R})} \, ]{}^t  
 }  ~~~,
 \label{DefGmm}
\ee
is utilized to form square matrices from any of the on-shell $ {\bm {\rL}}_\rI{}^{({\cal R})}$
quantities for all of the representations.  Given the form of these on-shell $ {\bm {\rL}}_\rI
{}^{({\cal R})}$ and $ {\bm {\rR}}_\rI{}^{({\cal R})}$ matrices,  some of the quantities ${\Hat 
{\bm \g}}{}_\rI{}^{({\cal R})}$ are elements of Coxeter Groups.  In any of the four on-shell 
representations, a quantity $ {\Hat {\bm {\cal C}}}{}^{({\cal R})}$ is defined as below 
from using each ${\Hat {\bm \g}}{}_\rI{}^{({\cal R})}$ representation, 
\begin{align}    
{\Hat {\bm {\cal C}}}^{({\cal R})} &= {\Hat {\bm \g}}_1^{({\cal R})}   \,  \cdots   \, 
{\Hat {\bm \g}}_4^{({\cal R})}  ~~~.
\end{align}
In the works \cite{HYMN1,HYMN2}, designed to quantify the relationships among adinkras 
related to one another by the raising or lowering of nodes, the eigenvalues of such 
operators are found to be the ``height-yielding matrix numbers" (HYMN) that specify this 
data in matrices associated with each supermultiplet.   However in the work of \cite{EX2} 
a different discovery was made.

The question in \cite{EX2} under investigation was whether matrices of the form of
${\Hat {\bm \g}}{}_\rI{}^{({\cal R})}$ and ${\Hat {\bm {\cal C}}}{}^{({\cal R})}$ carried 
the data to distinguish when pairs of minimal off-shell 4D, $\cal N$ = 1 supermultiplets 
could form off-shell 4D, $\cal N$ = 2 supermultiplets?  Here we ask the equivalent
question about going from 4D, $\cal N$ = 0 supermultiplets to 4D, $\cal N$ = 1 supermultiplets.

Next we calculate the anti-commutators of the matrices define in Eq. (\ref{DefGmm}).  
We find for ${({\cal R})}$ = $(TM)$, 
\begin{equation}
\left\{ \, {\Hat {\bm \g}}{}_\rI^{({TM})} ~,~ {\Hat {\bm \g}}{}_\rJ^{({TM})}  \, \right\} = 2 \,  
\delta_{\rm {IJ}} \, {\bm {\rm{I}}}{}_{8} 
~~~.
\label{CLFF2}
\end{equation}
However for the on-shell $(CM)$,  $(VM)$, and $(DTM)$ representations one can prove it
is impossible to satisfy an equation of this form.  Nor can we find \cite{pHEDRON}
\begin{equation}
\left\{ \, {\Hat {\bm \g}}{}_\rI ^{({\cal R})}~,~ {\Hat {\bm \g}}{}_\rJ^{({\cal R})}  \, \right\} = 
2\,  \delta_{\rm {IJ}} \, {\bm {\rm{I}}}{}_{d}  ~+~ {\cal N}{}_{\rm {IJ}}{}^{\Hat \a \, ({\cal 
R})} \, {\bm {\kappa}}{}_{\Hat \a}^{({\cal R})}
~~~,
\label{CLFF3}
\end{equation}
with a set of calculable $ {\cal N}{}_{\rm {IJ}}{}^{\Hat \a \, ({\cal R})} $ coefficients and 
d $\times$ d matrices ${\bm {\kappa}}{}_{\Hat \a}^{({\cal R})}$ \cite{adnkKyeoh}.
The non-invertability intrudes.

\section{Explicit Matrices Reported}
\label{sec:EgNRprt}

We now turn to the explicit representation of the matrices $ {\Hat {\bm \g}}{}_\rI ^{({\cal R})} $
and $ {\Hat {\bm {\cal C}}}^{({\cal R})}$ for the well-known cases of the chiral, tensor, and
vector supermultiplet, as the double tensor supermultiplet \cite{DZ1,DZ2,TV,BH}.

\subsection{On-shell Chiral Supermultiplet}
\begin{align*}
{\Hat {\bm \g}}{}_{1}^{(CM)}  &~=~ \left[
\begin{array}{cccccc}
 0 & 0 & 1 & 0 & 0 & 0 \\
 0 & 0 & 0 & 0 & 0 & -1 \\
 1 & 0 & 0 & 0 & 0 & 0 \\
 0 & 0 & 0 & 0 & 0 & 0 \\
 0 & 0 & 0 & 0 & 0 & 0 \\
 0 & -1 & 0 & 0 & 0 & 0 \\
\end{array} \right]  ~~~,~~~
 {\Hat {\bm \g}}{}_{2} ^{(CM)}  = \left[
\begin{array}{cccccc}
 0 & 0 & 0 & 1 & 0 & 0 \\
 0 & 0 & 0 & 0 & 1 & 0 \\
 0 & 0 & 0 & 0 & 0 & 0 \\
 1 & 0 & 0 & 0 & 0 & 0 \\
 0 & 1 & 0 & 0 & 0 & 0 \\
 0 & 0 & 0 & 0 & 0 & 0 \\
\end{array} \right]    ~~~,~~~ \\
{\Hat {\bm \g}}{}_{3} ^{(CM)}   &= \left[
\begin{array}{cccccc}
 0 & 0 & 0 & 0 & 1 & 0 \\
 0 & 0 & 0 & -1 & 0 & 0 \\
 0 & 0 & 0 & 0 & 0 & 0 \\
 0 & -1 & 0 & 0 & 0 & 0 \\
 1 & 0 & 0 & 0 & 0 & 0 \\
 0 & 0 & 0 & 0 & 0 & 0 \\
\end{array} \right]    ~~~,~~~
{\Hat {\bm \g}}{}_{4} ^{(CM)}  = \left[
\begin{array}{cccccc}
 0 & 0 & 0 & 0 & 0 & 1 \\
 0 & 0 & 1 & 0 & 0 & 0 \\
 0 & 1 & 0 & 0 & 0 & 0 \\
 0 & 0 & 0 & 0 & 0 & 0 \\
 0 & 0 & 0 & 0 & 0 & 0 \\
 1 & 0 & 0 & 0 & 0 & 0 \\
\end{array}
\right]  ~~~, \\
\Hat{\mathcal{C}} &=\left[
\begin{array}{cccccc}
 0 & 0 & 0 & 0 & 0 & 0 \\
 0 & 0 & 0 & 0 & 0 & 0 \\
 0 & 0 & -1 & 0 & 0 & 0 \\
 0 & 0 & 0 & 0 & 0 & 0 \\
 0 & 0 & 0 & 0 & 0 & 0 \\
 0 & 0 & 0 & 0 & 0 & -1 \\
\end{array}
\right] ~~~.
\end{align*}

\subsection{On-shell Tensor Supermultiplet}
\begin{align*}
{\Hat {\bm \g}}{}_{1}^{(TM)} &= \left[
\begin{array}{cccccccc}
 0 & 0 & 0 & 0 & 1 & 0 & 0 & 0 \\
 0 & 0 & 0 & 0 & 0 & 0 & -1 & 0 \\
 0 & 0 & 0 & 0 & 0 & 0 & 0 & -1 \\
 0 & 0 & 0 & 0 & 0 & -1 & 0 & 0 \\
 1 & 0 & 0 & 0 & 0 & 0 & 0 & 0 \\
 0 & 0 & 0 & -1 & 0 & 0 & 0 & 0 \\
 0 & -1 & 0 & 0 & 0 & 0 & 0 & 0 \\
 0 & 0 & -1 & 0 & 0 & 0 & 0 & 0 \\
\end{array} \right]   ~~~,~~~
{\Hat {\bm \g}}{}_{2}^{(TM)}    = \left[
\begin{array}{cccccccc}
 0 & 0 & 0 & 0 & 0 & 1 & 0 & 0 \\
 0 & 0 & 0 & 0 & 0 & 0 & 0 & 1 \\
 0 & 0 & 0 & 0 & 0 & 0 & -1 & 0 \\
 0 & 0 & 0 & 0 & 1 & 0 & 0 & 0 \\
 0 & 0 & 0 & 1 & 0 & 0 & 0 & 0 \\
 1 & 0 & 0 & 0 & 0 & 0 & 0 & 0 \\
 0 & 0 & -1 & 0 & 0 & 0 & 0 & 0 \\
 0 & 1 & 0 & 0 & 0 & 0 & 0 & 0 \\
\end{array}
\right]  ~~~,  \\
{\Hat {\bm \g}}{}_{3}^{(TM)} &= \left[
\begin{array}{cccccccc}
 0 & 0 & 0 & 0 & 0 & 0 & 1 & 0 \\
 0 & 0 & 0 & 0 & 1 & 0 & 0 & 0 \\
 0 & 0 & 0 & 0 & 0 & 1 & 0 & 0 \\
 0 & 0 & 0 & 0 & 0 & 0 & 0 & -1 \\
 0 & 1 & 0 & 0 & 0 & 0 & 0 & 0 \\
 0 & 0 & 1 & 0 & 0 & 0 & 0 & 0 \\
 1 & 0 & 0 & 0 & 0 & 0 & 0 & 0 \\
 0 & 0 & 0 & -1 & 0 & 0 & 0 & 0 \\
\end{array}
\right]   ~~~,~~~
{\Hat {\bm \g}}{}_{4}^{(TM)} = \left[
\begin{array}{cccccccc}
 0 & 0 & 0 & 0 & 0 & 0 & 0 & 1 \\
 0 & 0 & 0 & 0 & 0 & -1 & 0 & 0 \\
 0 & 0 & 0 & 0 & 1 & 0 & 0 & 0 \\
 0 & 0 & 0 & 0 & 0 & 0 & 1 & 0 \\
 0 & 0 & 1 & 0 & 0 & 0 & 0 & 0 \\
 0 & -1 & 0 & 0 & 0 & 0 & 0 & 0 \\
 0 & 0 & 0 & 1 & 0 & 0 & 0 & 0 \\
 1 & 0 & 0 & 0 & 0 & 0 & 0 & 0 \\
\end{array}
\right] ~~~, \\
\Hat{\mathcal{C}}{}^{(TM)} &=\left[
\begin{array}{cccccccc}
 -1 & 0 & 0 & 0 & 0 & 0 & 0 & 0 \\
 0 & -1 & 0 & 0 & 0 & 0 & 0 & 0 \\
 0 & 0 & -1 & 0 & 0 & 0 & 0 & 0 \\
 0 & 0 & 0 & -1 & 0 & 0 & 0 & 0 \\
 0 & 0 & 0 & 0 & 1 & 0 & 0 & 0 \\
 0 & 0 & 0 & 0 & 0 & 1 & 0 & 0 \\
 0 & 0 & 0 & 0 & 0 & 0 & 1 & 0 \\
 0 & 0 & 0 & 0 & 0 & 0 & 0 & 1 \\
\end{array}
\right]  ~~~. 
\end{align*}

\subsection{On-shell Vector Supermultiplet}
\begin{align*}
{\Hat {\bm \g}}{}_{1}^{(VM)} &= \left[
\begin{array}{ccccccc}
 0 & 0 & 0 & 0 & 1 & 0 & 0 \\
 0 & 0 & 0 & 0 & 0 & 0 & -1 \\
 0 & 0 & 0 & 1 & 0 & 0 & 0 \\
 0 & 0 & 1 & 0 & 0 & 0 & 0 \\
 1 & 0 & 0 & 0 & 0 & 0 & 0 \\
 0 & 0 & 0 & 0 & 0 & 0 & 0 \\
 0 & -1 & 0 & 0 & 0 & 0 & 0 \\
\end{array}
\right]   ~~~,~~~
{\Hat {\bm \g}}{}_{2}^{(VM)} = \left[
\begin{array}{ccccccc}
 0 & 0 & 0 & 1 & 0 & 0 & 0 \\
 0 & 0 & 0 & 0 & 0 & 1 & 0 \\
 0 & 0 & 0 & 0 & -1 & 0 & 0 \\
 1 & 0 & 0 & 0 & 0 & 0 & 0 \\
 0 & 0 & -1 & 0 & 0 & 0 & 0 \\
 0 & 1 & 0 & 0 & 0 & 0 & 0 \\
 0 & 0 & 0 & 0 & 0 & 0 & 0 \\
\end{array}
\right]  ~~~,  \\
{\Hat {\bm \g}}{}_{3}^{(VM)}
&= \left[
\begin{array}{ccccccc}
 0 & 0 & 0 & 0 & 0 & 0 & 1 \\
 0 & 0 & 0 & 0 & 1 & 0 & 0 \\
 0 & 0 & 0 & 0 & 0 & 1 & 0 \\
 0 & 0 & 0 & 0 & 0 & 0 & 0 \\
 0 & 1 & 0 & 0 & 0 & 0 & 0 \\
 0 & 0 & 1 & 0 & 0 & 0 & 0 \\
 1 & 0 & 0 & 0 & 0 & 0 & 0 \\
\end{array}
\right]   ~~~,~~~
{\Hat {\bm \g}}{}_{4}^{(VM)} = \left[
\begin{array}{ccccccc}
 0 & 0 & 0 & 0 & 0 & 1 & 0 \\
 0 & 0 & 0 & -1 & 0 & 0 & 0 \\
 0 & 0 & 0 & 0 & 0 & 0 & -1 \\
 0 & -1 & 0 & 0 & 0 & 0 & 0 \\
 0 & 0 & 0 & 0 & 0 & 0 & 0 \\
 1 & 0 & 0 & 0 & 0 & 0 & 0 \\
 0 & 0 & -1 & 0 & 0 & 0 & 0 \\
\end{array}
\right]  ~~~,  \\
\Hat{\mathcal{C}}{}^{(VM)}   &=\left[
\begin{array}{ccccccc}
 -1 & 0 & 0 & 0 & 0 & 0 & 0 \\
 0 & 0 & 0 & 0 & 0 & 0 & 0 \\
 0 & 0 & -1 & 0 & 0 & 0 & 0 \\
 0 & 0 & 0 & 1 & 0 & 0 & 0 \\
 0 & 0 & 0 & 0 & 0 & 0 & 0 \\
 0 & 0 & 0 & 0 & 0 & 0 & 0 \\
 0 & 0 & 0 & 0 & 0 & 0 & 1 \\
\end{array}
\right] ~~~.
\end{align*}

\subsection{On-shell Double Tensor Supermultiplet} 
\begin{align*}
{\Hat {\bm \g}}{}_{1}^{(DTM)} &= \left[
\begin{array}{cccccccccc}
 0 & 0 & 0 & 0 & 0 & 0 & 0 & -1 & 0 & 0 \\
 0 & 0 & 0 & 0 & 0 & 0 & 1 & 0 & 0 & 0 \\
 0 & 0 & 0 & 0 & 0 & 0 & 0 & 0 & 1 & 0 \\
 0 & 0 & 0 & 0 & 0 & 0 & 0 & 0 & -1 & 0 \\
 0 & 0 & 0 & 0 & 0 & 0 & 0 & 0 & 0 & -1 \\
 0 & 0 & 0 & 0 & 0 & 0 & 0 & -1 & 0 & 0 \\
 0 & 1 & 0 & 0 & 0 & 0 & 0 & 0 & 0 & 0 \\
 -1 & 0 & 0 & 0 & 0 & -1 & 0 & 0 & 0 & 0 \\
 0 & 0 & 1 & -1 & 0 & 0 & 0 & 0 & 0 & 0 \\
 0 & 0 & 0 & 0 & -1 & 0 & 0 & 0 & 0 & 0 \\
\end{array}
\right]    ~~~,~~~ \\
{\Hat {\bm \g}}{}_{2}^{(DTM)} &= \left[
\begin{array}{cccccccccc}
 0 & 0 & 0 & 0 & 0 & 0 & -1 & 0 & 0 & 0 \\
 0 & 0 & 0 & 0 & 0 & 0 & 0 & -1 & 0 & 0 \\
 0 & 0 & 0 & 0 & 0 & 0 & 0 & 0 & 0 & 1 \\
 0 & 0 & 0 & 0 & 0 & 0 & 0 & 0 & 0 & 1 \\
 0 & 0 & 0 & 0 & 0 & 0 & 0 & 0 & -1 & 0 \\
 0 & 0 & 0 & 0 & 0 & 0 & 1 & 0 & 0 & 0 \\
 -1 & 0 & 0 & 0 & 0 & 1 & 0 & 0 & 0 & 0 \\
 0 & -1 & 0 & 0 & 0 & 0 & 0 & 0 & 0 & 0 \\
 0 & 0 & 0 & 0 & -1 & 0 & 0 & 0 & 0 & 0 \\
 0 & 0 & 1 & 1 & 0 & 0 & 0 & 0 & 0 & 0 \\
\end{array}
\right]    ~~~,  \\
{\Hat {\bm \g}}{}_{3}^{(DTM)} &= \left[
\begin{array}{cccccccccc}
 0 & 0 & 0 & 0 & 0 & 0 & 0 & 0 & 0 & 1 \\
 0 & 0 & 0 & 0 & 0 & 0 & 0 & 0 & -1 & 0 \\
 0 & 0 & 0 & 0 & 0 & 0 & 1 & 0 & 0 & 0 \\
 0 & 0 & 0 & 0 & 0 & 0 & 1 & 0 & 0 & 0 \\
 0 & 0 & 0 & 0 & 0 & 0 & 0 & 1 & 0 & 0 \\
 0 & 0 & 0 & 0 & 0 & 0 & 0 & 0 & 0 & -1 \\
 0 & 0 & 1 & 1 & 0 & 0 & 0 & 0 & 0 & 0 \\
 0 & 0 & 0 & 0 & 1 & 0 & 0 & 0 & 0 & 0 \\
 0 & -1 & 0 & 0 & 0 & 0 & 0 & 0 & 0 & 0 \\
 1 & 0 & 0 & 0 & 0 & -1 & 0 & 0 & 0 & 0 \\
\end{array}
\right]  ~~~,  \\
{\Hat {\bm \g}}{}_{4}^{(DTM)} &= \left[
\begin{array}{cccccccccc}
 0 & 0 & 0 & 0 & 0 & 0 & 0 & 0 & 1 & 0 \\
 0 & 0 & 0 & 0 & 0 & 0 & 0 & 0 & 0 & 1 \\
 0 & 0 & 0 & 0 & 0 & 0 & 0 & 1 & 1 & 0 \\
 0 & 0 & 0 & 0 & 0 & 0 & 0 & -1 & -1 & 0 \\
 0 & 0 & 0 & 0 & 0 & 0 & 0 & 0 & 0 & 0 \\
 0 & 0 & 0 & 0 & 0 & 0 & 0 & 0 & 1 & 0 \\
 0 & 0 & 0 & 0 & 0 & 0 & 0 & 0 & 0 & 0 \\
 0 & 0 & 1 & -1 & 0 & 0 & 0 & 0 & 0 & 0 \\
 1 & 0 & 1 & -1 & 0 & 1 & 0 & 0 & 0 & 0 \\
 0 & 1 & 0 & 0 & 0 & 0 & 0 & 0 & 0 & 0 \\
\end{array}
\right]  ~~~, \\
\Hat{\mathcal{C}}{}^{(DTM)} &=\left[
\begin{array}{cccccccccc}
 -1 & 0 & -1 & 1 & 0 & -1 & 0 & 0 & 0 & 0 \\
 0 & -2 & 0 & 0 & 0 & 0 & 0 & 0 & 0 & 0 \\
 0 & 0 & -1 & 1 & 0 & 0 & 0 & 0 & 0 & 0 \\
 0 & 0 & 1 & -1 & 0 & 0 & 0 & 0 & 0 & 0 \\
 0 & 0 & 0 & 0 & 0 & 0 & 0 & 0 & 0 & 0 \\
 -1 & 0 & -1 & 1 & 0 & -1 & 0 & 0 & 0 & 0 \\
 0 & 0 & 0 & 0 & 0 & 0 & 0 & 0 & 0 & 0 \\
 0 & 0 & 0 & 0 & 0 & 0 & 0 & 0 & 0 & 0 \\
 0 & 0 & 0 & 0 & 0 & 0 & 0 & 0 & 0 & 0 \\
 0 & 0 & 0 & 0 & 0 & 0 & 0 & 0 & 0 & -1 \\
\end{array}
\right]  ~~~.
\end{align*}
This final result is so distinctly different from the other cases,
reinforing the description of $(DTM)$ in \cite{G-1} as ``pathogenic.''   The eigenvalues of
$ \Hat{\mathcal{C}}{}^{(DTM)}$ are $\{-2, -2, -2, -1, 0, 0,  0, 0,  0,  0 \}$.  In no other 
case does a factor of 2 appear.  In all the other cases, $ \Hat{\mathcal{C}}{}^{(R)}$ 
(for ${({\cal R})} \in \{ {(CM)}, {(TM)}, {(VM)} \}$) is diagonal and with 
augmentation as described in \cite{McInT} become invertible.  This suggest there
may be added difficulties in achieving a full off-shell version of the 4D, $\cal N $ =
 1 $(DTM)$.

\vspace{.05in}
\begin{center}
\parbox{4in}{{\it ``An ocean traveller has even more vividly the impression  \\ $~~$ 
that the ocean is made of waves than that it is made of \\ $~~$ 
water.' \\ ${~}$ 
${~}$ 
\\ ${~}$ }\,\,-\,
Arthur Eddington $~~~~~~~~~$}
\parbox{4in}{
$~~$}  
\end{center}
		
\noindent
{\bf Acknowledgments}\\[.1in] \indent
D.\ G.\ acknowledges support from NSF GRFP Grant No.\ DGE1144152 and 
NSF Grant No.\ 17079.  We wish also to acknowledge helpful conversation 
with M.\ Faux, M.\ Ro\v cek, and W.\ Siegel.  The research of SJG is supported 
by the endowment of the Ford Foundation Professorship of Physics at Brown 
University and the Brown Theoretical Physics Center.

\end{document}